\documentstyle[camera,prbplug,psfig]{article}

\begin{document}

\title{ {\small Journal of Physics: Condensed Matter 14 (2002) 21}\\
\vspace{0.5cm}
Stripe formation in high-$T_c$ superconductors 
}

\author{Takashi Yanagisawa, Soh Koike, Mitake Miyazaki, and Kunihiko Yamaji}

\address{Condensed-Matter Physics Group, 
Nanoelectronics Research Institute,
AIST Tsukuba Central 2,
1-1-1 Umezono, Tsukuba,
Ibaraki 305-8568, Japan}

\date{(3 August 2001)}

\maketitle

\begin{abstract}
The non-uniform ground state of the two-dimensional three-band
Hubbard model for the oxide high-$T_c$ superconductors is investigated
using a variational Mont Carlo method. 
We examine the effect produced by holes doped into the antiferromagnetic (AF) 
background in the underdoped region.
It is shown that the AF state with spin modulations and
stripes is stabilized due to holes travelling in the CuO plane.
The structures of modulated AF spins are dependent upon the 
parameters used in the model.
The effect of boundary conditions is reduced for large systems.
We show that there is a region where incommensurability is proportional 
to the hole density.
Our results give a consistent description of stripes observed by the neutron
scattering experiments based on the three-band model for the CuO plane.
\\
\end{abstract}

 
\section{Introduction}
A mechanism of superconductivity of high-$T_c$ cuprates is not still
clarified after the intensive efforts over a decade.  An origin of the 
anomalous metallic properties in the underdoped region has also been 
investigated by many physicists as a challenging problem.
In order to solve the mysteries of high-$T_c$ cuprates,
it is important to examine the ground state of the
two-dimensional CuO$_2$ planes which are usually contained
in the crystal structures of high-$T_c$ oxide superconductors.\cite{lt99}
A basic model for the CuO$_2$ plane is the two-dimensional three-band Hubbard 
model with 
$d$ and $p$ orbitals, which is expected to contain essential features of
high-$T_c$ cuprates.\cite{eme87,tje89}  
The undoped oxide compounds exhibit a rich structure of 
antiferromagnetic (AF) correlations over a wide range of temperature described by
the two-dimensional quantum 
antiferromagnetism.\cite{shi87,lyo88,aep88,man89,din90}
It is also considered that a small number of holes introduced by doping
are responsible for the disappearance of long-range AF
ordering.\cite{zha88,pre88,inu88,yan92}
Recent
neutron-scattering experiments have suggested an existence of incommensurate
ground states with modulation vectors given by $Q_s=(\pi\pm 2\pi\delta,\pi)$
and $Q_c=(\pm4\pi\delta,0)$ (or $Q_s=(\pi,\pi\pm2\pi\delta)$ and 
$Q_c=(0,\pm 4\pi\delta)$ where $\delta$ denotes the hole-doping 
ratio.\cite{tra95}
We can expect that the incommensurate correlations are induced by holes
moving around in the Cu-O plane in the underdoped region.

The purpose of this paper is to investigate the effect of hole doping in the
ground state of the three-band Hubbard model in the underdoped region
using a variational Monte Carlo method\cite{nak97,yam98,yan98} which is a
tool to control the correlation from weakly to strongly correlated regions.
It is shown that the AF long-range ordering disappears due
to extra holes doped into the two-dimensional plane.
With respect to the initial indications given by the neutron-scattering 
measurements,
the possibility of incommensurate stripe states is examined concerning any 
dependences on the hole density $\delta$, especially regarding the region 
near of 1/8 
doping.  Although the possible incommensurate states are sensitively
dependent upon the boundary conditions in small systems, the effect
of boundary conditions is reduced for larger systems.

The paper is arranged as follows.  In Section II the wave functions and
the method for the three-band Hubbard model are described.  In Section III
the results are shown and the last Section summarizes the study.

\section{2D three-band Hubbard model and wave functions}
The three-band Hubbard model has been investigated intensively with respect to
superconductivity (SC) in cuprate high-$T_c$ 
materials.\cite{oga88,ste89,hir89,sca91,dop90,dop92,hot94,asa96,kur96,tak97,gue98,koi00,yan00,koi01}
However,
a non-uniform AF ground state for the three-band model has
not yet been examined as intensively.\cite{zaa96}
The three-band Hubbard model is written as\cite{hir89,yan01,yan01b}
\begin{eqnarray}
H&=& \epsilon_d\sum_{i\sigma}d^{\dag}_{i\sigma}d_{i\sigma} + 
U\sum_id^{\dag}_{i\uparrow}d_{i\uparrow}d^{\dag}_{i\downarrow}d_{i\downarrow}
\nonumber\\
&+& \epsilon_p\sum_{i\sigma}(p^{\dag}_{i+\hat{x}/2,\sigma}p_{i+\hat{x}/2,\sigma}
+p^{\dag}_{i+\hat{y}/2,\sigma}p_{i+\hat{y}/2,\sigma})\nonumber\\
&+& t_{dp}\sum_{i\sigma}[d^{\dag}_{i\sigma}(p_{i+\hat{x}/2,\sigma}
+p_{i+\hat{y}/2,\sigma}-p_{i-\hat{x}/2,\sigma}\nonumber\\
&-& p_{i-\hat{y}/2,\sigma})+h.c.]\nonumber\\
&+& t_{pp}\sum_{i\sigma}[p^{\dag}_{i+\hat{y}/2,\sigma}p_{i+\hat{x}/2,\sigma}
-p^{\dag}_{i+\hat{y}/2,\sigma}p_{i-\hat{x}/2,\sigma}\nonumber\\
&-&p^{\dag}_{i-\hat{y}/2,\sigma}p_{i+\hat{x}/2,\sigma}
+p^{\dag}_{i-\hat{y}/2,\sigma}p_{i-\hat{x}/2,\sigma} +h.c.].\nonumber\\
\end{eqnarray}
$\hat{x}$ and $\hat{y}$ represent unit vectors in the x and y directions,
respectively,
$p^{\dag}_{i\pm\hat{x}/2,\sigma}$
and $p_{i\pm\hat{x}/2,\sigma}$ denote the operators for the $p$ electrons at
the site $R_i\pm\hat{x}/2$, and in a similar way $p^{\dag}_{i\pm\hat{y}/2,\sigma}$ 
and $p_{i\pm\hat{y}/2,\sigma}$ are  defined.
$U(\equiv U_d)$ denotes the strength of Coulomb interaction between the $d$ 
electrons.
For simplicity we neglect the Coulomb interaction among $p$ electrons.
Other notations are standard and energies are measured in 
$t_{dp}$ units.
The number of cells which consist of $d$, $p_x$ and $p_y$ orbitals is 
denoted as $N$.

The wave functions are given by the 
normal state, spin density wave (SDW) and modulated SDW wave functions with the 
Gutzwiller projection.
For the three-band Hubbard model the wave functions for normal and SDW
states are written as
\begin{equation}
\psi_n= P_G\prod_{|k|\leq k_F,\sigma}\alpha_{k\sigma}^{\dag}|0\rangle ,
\end{equation}
\begin{equation}
\psi_{SDW}= P_G\prod_{|k|\leq k_F,\sigma}\beta_{k\sigma}^{\dag}|0\rangle ,
\end{equation}
where $\alpha_{k\sigma}$ is the linear combination of $d_{k\sigma}$,
$p_{xk\sigma}$ and $p_{yk\sigma}$ constructed to express an operator for the 
lowest band of a non-interacting Hamiltonian in the hole picture.  
$P_G$ is the Gutzwiller operator given by
\begin{equation}
P_G=\prod_i(1-(1-g)n_{di\uparrow}n_{di\downarrow}),
\end{equation}
for $n_{di\sigma}=d^{\dag}_{i\sigma}d_{i\sigma}$.
For $t_{pp}=0$,
$\alpha_{k\sigma}$ is expressed in terms of a variational parameter
$\tilde{\epsilon_p}-\tilde{\epsilon_d}$ as follows:
\onecolumn
\begin{equation}
\alpha^{\dag}_{k\sigma}= \left(\frac{1}{2}\left(
1+\frac{\tilde{\epsilon_p}-\tilde{\epsilon_d}}{2E_k}\right)\right)^{1/2}d^{\dag}_{k\sigma}
+{\rm i}\left(\frac{1}{2}\left(
1-\frac{\tilde{\epsilon_p}-\tilde{\epsilon_d}}{2E_k}\right)\right)^{1/2}\left(
\frac{w_{xk}}{w_k}p^{\dag}_{xk\sigma}
+\frac{w_{yk}}{w_k}p^{\dag}_{yk\sigma}\right),
\end{equation}
\twocolumn
where $w_{xk}=2t_{dp}{\rm sin}(k_x/2)$, $w_{yk}=2t_{dp}{\rm sin}(k_y/2)$,
$w_k=(w_{xk}^2+w_{yk}^2)^{1/2}$ and 
$E_k=[(\tilde{\epsilon_p}-\tilde{\epsilon_d})^2/4+w_k^2]^{1/2}$.
For the commensurate SDW $\beta_{k\sigma}$ is given by a linear
combination of $d_{k\sigma}$, $p_{xk\sigma}$, $p_{yk\sigma}$, $d_{k+Q\sigma}$, 
$p_{xk+Q\sigma}$ and $p_{yk+Q\sigma}$ for $Q=(\pi,\pi)$.
$P_G$ is the Gutzwiller projection operator for the Cu $d$ site.
We can easily generalize it to the incommensurate case by diagonalizing the
Hartree-Fock Hamiltonian.
The wave function with a stripe can be taken to be Gutzwiller, i.e.
\begin{equation}
\psi_{stripe}=P_G\psi_{stripe}^0.
\end{equation}
$\psi_{stripe}^0$ is the Slater determinant made from solutions of
the Hartree-Fock Hamiltonian given as
\begin{equation}
H_{trial}=
H^0_{dp}+\sum_{i\sigma}[\delta n_{di}-\sigma(-1)^{x_i+y_i}m_i]
d^{\dag}_{i\sigma}d_{i\sigma},
\end{equation}
where $H^0_{dp}$ is the non-interacting part of the Hamiltonian $H$ with
the variational parameter $\tilde{\epsilon_p}$ and $\tilde{\epsilon_d}$.
The Slater determinant is constructed from wave functions of $N_e/2$ lowest
eigenstates after diagonalizing $H_{trial}$ in $k$-space for each spin, 
where $N_e$ is the number of electrons.
$\delta n_{di}$ and $m_i$ are expressed by modulation vectors $Q_s$ and $Q_c$ 
reprsenting the spin and charge part, respectively.
In this paper $\delta n_{di}$ and $m_i$ are assumed to have the 
form\cite{yan01b,gia91} 
\begin{equation}
\delta n_{di}=-\sum_j \alpha/{\rm cosh}((x_i-x^{str}_j)/\xi_c),
\label{ndi}
\end{equation}
\begin{equation}
m_i=\Delta_{incom}\prod_j {\rm tanh}((x_i-x^{str}_j)/\xi_s),
\label{mi}
\end{equation}
with parameters $\alpha$, $\Delta_{incom}$, $\xi_c$ and $\xi_s$, where 
$x^{str}_j$ denote the position of a stripe.

A Monte Carlo algorithm developed using the auxiliary-field quantum Monte Carlo
calculations is employed to evaluate the expected values for the wave
functions shown above.\cite{yan98,bla81}
Using the discrete Hubbard-Stratonovich transformation, the Gutzwiller
factor is written as
\begin{equation}
{\rm exp}(-\alpha\sum_i n_{di\uparrow}n_{di\downarrow})=
(\frac{1}{2})^N \sum_{\{s_i\}}{\rm exp}[2a\sum_i s_i(n_{di\uparrow}-n_{di\downarrow})-\frac{\alpha}{2}\sum_i (n_{di\uparrow}+n_{di\downarrow})],
\end{equation}
where $\alpha={\rm log}(1/g)$ and ${\rm cosh}(2a)={\rm e}^{\alpha/2}$.
The Hubbard-Stratonovich auxiliary field $s_i$ takes the values of $\pm 1$.
The norm $\langle\psi_{stripe}|\psi_{stripe}\rangle$ is written as
\onecolumn
\begin{equation}
\langle\psi_{stripe}|\psi_{stripe}\rangle={\rm const.}\sum_{\{u_i\}\{s_i\}}
\prod_{\sigma}{\rm det}(\phi^{\sigma\dag}_0{\rm exp}(V^{\sigma}(u,\alpha))
{\rm exp}(V^{\sigma}(s,\alpha))\phi^{\sigma}_0),
\end{equation}
\twocolumn
where $V^{\sigma}(s,\alpha)$ is a diagonal $3N\times 3N$ matrix corresponding
to the potential
\begin{equation}
h^{\sigma}(s)= 2a\sigma\sum_i s_in_{di\sigma}-\frac{\alpha}{2}\sum_i
n_{di\sigma}.
\end{equation}
$V^{\sigma}(s,\alpha)$ is given by
$V^{\sigma}(s,\alpha)= {\rm diag}(2a\sigma s_1-\alpha/2,\cdots,2a\sigma s_N-\alpha/2,0,\cdots)$ where diag($a,\cdots$) denotes a diagonal matrix with its elements
given by the arguments $a,\cdots$.  $V^{\sigma}(s,\alpha)$ has non-zero
elements only for the $d$-electron part.
The elements of $(\phi^{\sigma}_0)_{ij}$ ($i=1,\cdots,3N;j=1,\cdots,N_e/2$) 
are given by linear combinations of plane waves:
\begin{equation}
(\phi^{\sigma}_0)_{ij}=\sum_{\ell}{\rm exp}(i{\bf r}_i\cdot
{\bf k}_{\ell})w^d_{\ell j} ,
\end{equation}
for $d$-electron part ($i=1,\cdots,N$) where $w^d_{\ell j}$ is the weight of
$d$ electrons for $\ell$-th wave vector and j-th lowest level from below obtained
from the diagonalization of $H_{trial}$.  The $p$-electron parts are similarly
defined.  Thus
\begin{equation}
(\phi^{\sigma}_0)_{ij}=\sum_{\ell}{\rm exp}(i{\bf r}_i\cdot
{\bf k}_{\ell})w^x_{\ell j}~~(i=N+1,\cdots,2N,j=1,\cdots,N_e/2) , 
\end{equation}
\begin{equation}
(\phi^{\sigma}_0)_{ij}=\sum_{\ell}{\rm exp}(i{\bf r}_i\cdot
{\bf k}_{\ell})w^y_{\ell j}~~(i=2N+1,\cdots,3N,j=1,\cdots,N_e/2) , 
\end{equation}
where $w^x_{\ell j}$ and $w^y_{\ell j}$ denote the weight of $p_x$ and
$p_y$ electrons, respectively.
Then we can apply the standard Monte Carlo sampling method to
evaluate the expectation values.\cite{bla81,yan98}
In order to perform a search for optimized values of the parameters included in the
wave functions, we employ a correlated-measurements method to reduce the
cpu time needed to find the most descendent direction in the parameter
space.\cite{umr88} 
In one Monte Carlo step all the Hubbard-Stratonovich variables are updated
once following the Metropolis algorithm.  We perform several $5\times 10^4$
Monte Carlo steps to evaluate the expectation values for optimized parameters.

\begin{figure}
\centerline{\psfig{figure=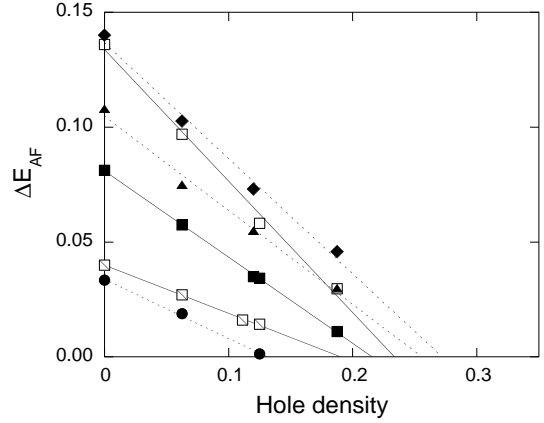,width=9cm}}
\caption{
Uniform SDW energy gain per site with reference to the normal-state energy as a 
function of the hole density $\delta$.
Data are from $8\times 8$, $10\times 10$, $12\times 12$ and $16\times 12$ 
systems for $\epsilon_p-\epsilon_d=2$. 
For solid symbols $U=4$ (circles), $U=8$ (squares), $U=12$ (triangles) and
$U=20$ (diamonds) for $t_{pp}=0.2$.
For open squares $U=8$ and $t_{pp}=0$ and for open squares with slash $U=8$
and $t_{pp}=0.4$.
The lines are a guide to eyes.  The Monte Carlo statistical errors are smaller than
the size of symbols.
}
\label{fig1}
\end{figure}

\begin{figure}
\centerline{\psfig{figure=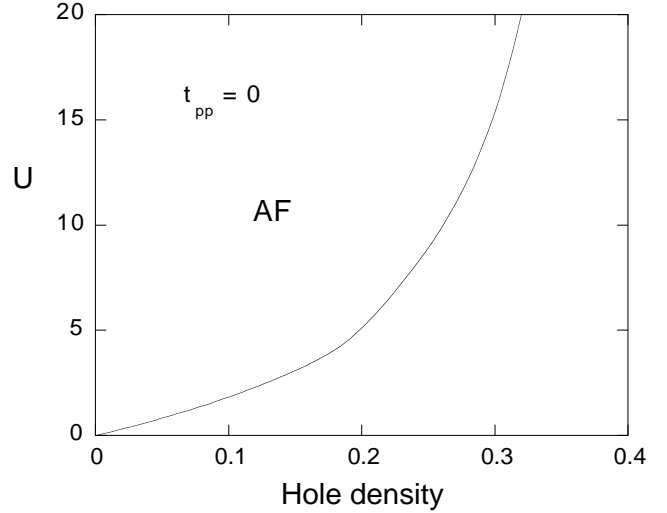,width=\columnwidth}}
\caption{
AF region in the plane of $U$ and the hole density for
$t_{pp}=0.2$ and $\epsilon_p-\epsilon_d=2$.  
}
\label{fig2}
\end{figure}

\begin{figure}
\centerline{\psfig{figure=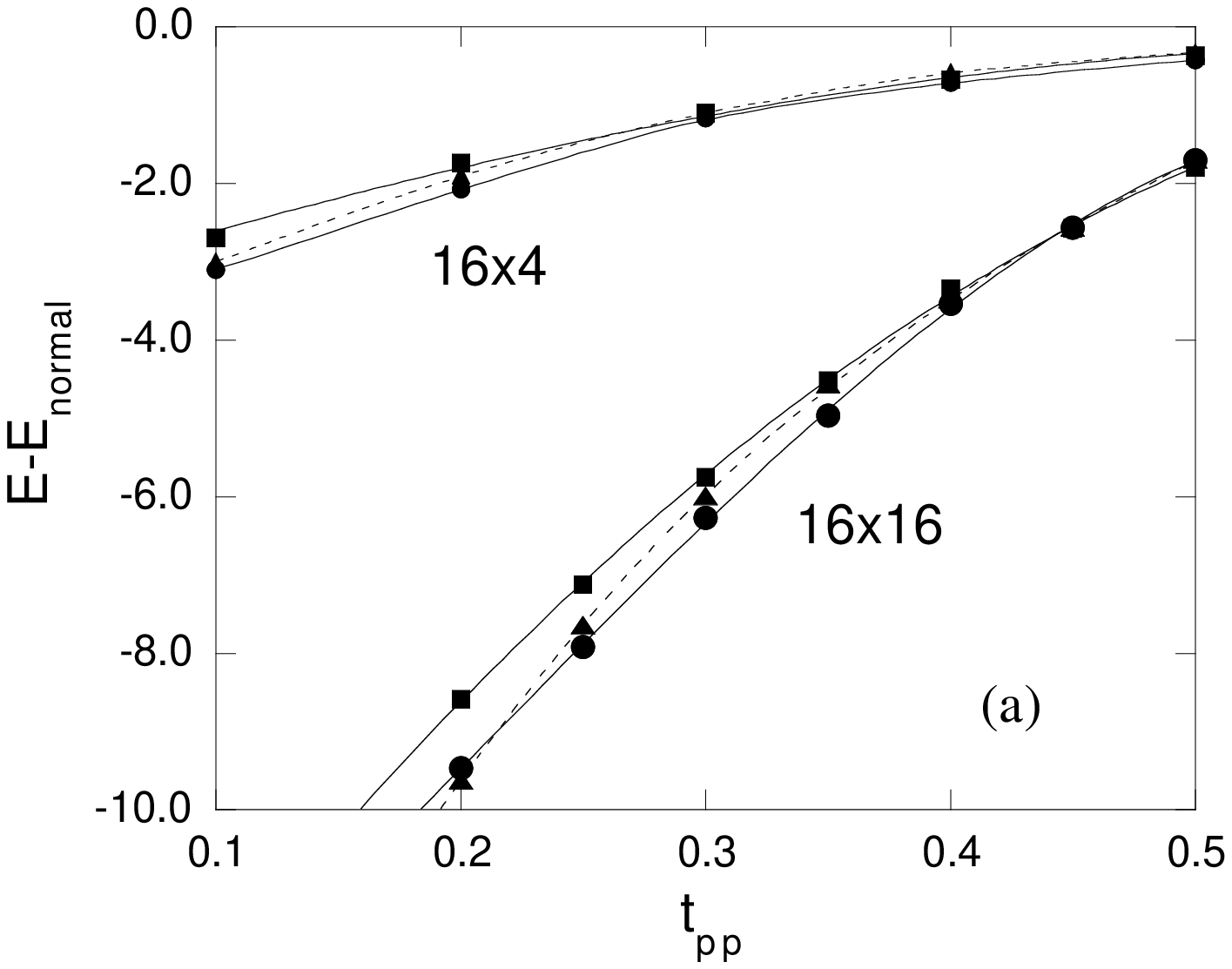,width=12cm}}
\centerline{\psfig{figure=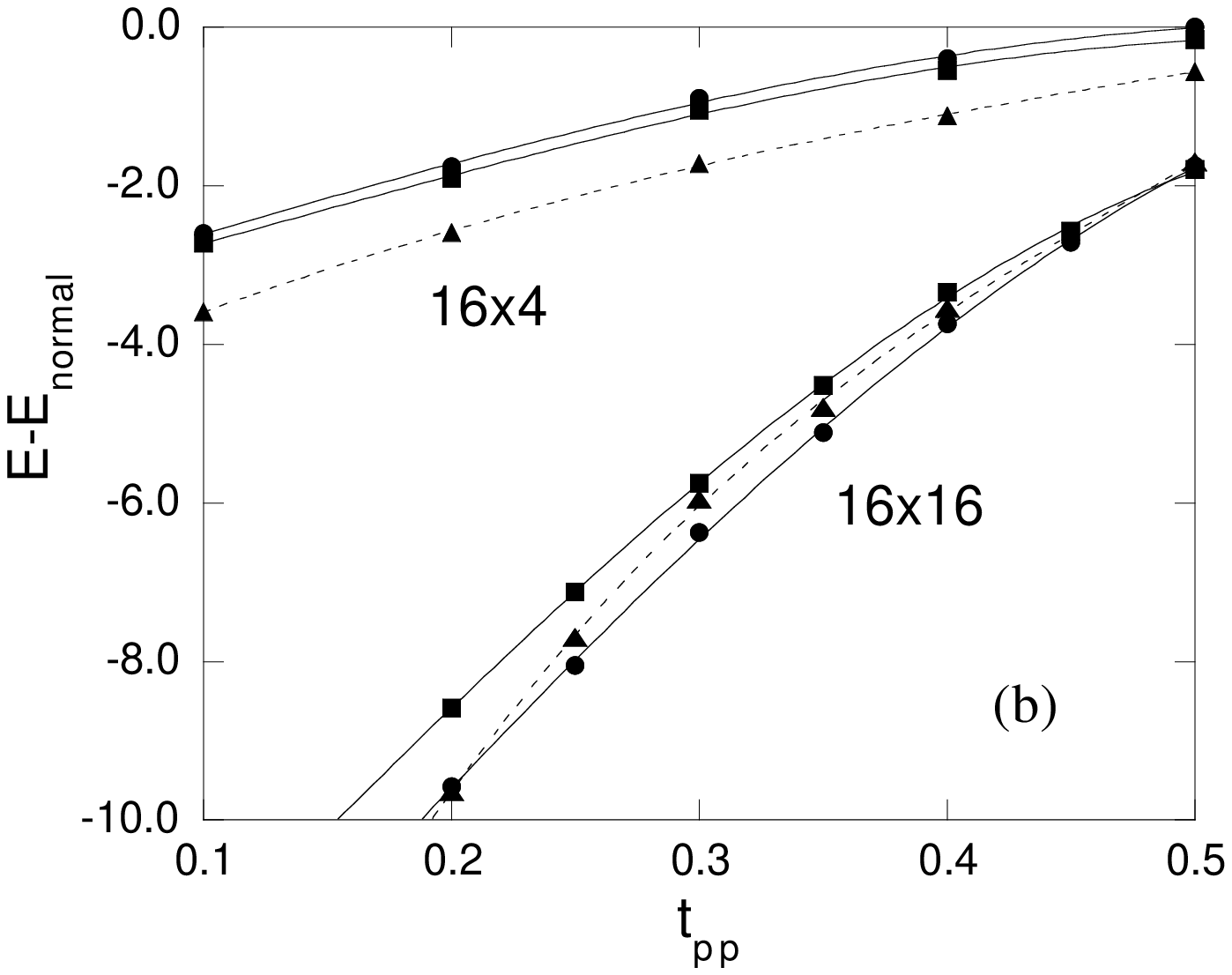,width=12cm}}
\caption{
Energy per site in reference to the normal state as a function of $t_{pp}$ for 
$16\times4$ and $16\times 16$ lattices at $\delta=1/8$.
Circles, triangles and squares denote the energy for 4-lattice stripes,
8-lattice stripes, and commensurate SDW, respectively, where
$n$-lattice stripe is the incommensurate state with one stripe per $n$
ladders.
In (a) the boundary conditions are antiperiodic in $x$-direction and
periodic in $y$-direction, and in (b) they are periodic in $x$-direction and
antiperiodic in $y$-direction.
The Monte Carlo statistical errors are within the size of symbols.
}
\label{fig3}
\end{figure}

\section{Antiferromagnetism and Stripes in the underdoped region}
We show the energy gain $\Delta E_{AF}$ for the uniform SDW state in reference
to the normal state for optimized parameters $g$, 
$\tilde{\epsilon_p}-\tilde{\epsilon_d}$ and AF order parameter
$\Delta_{AF}$ in Fig.1.
The energy is lowered considerably by the AF
long-range ordering up to about 20
of $U\approx 8-12$.

$\Delta E_{AF}$ decreases monotonically as $t_{pp}$ increases and
increases as $U$ increases.  One should note that $\Delta E_{AF}$ is larger
than the energy gain for the $d$-wave pairing state in the low-doping
region near the doping ratio $\delta\sim 0.1$ by two order of 
magnitude.\cite{yan00}
The boundary of the AF state in the plane of $U$ and the hole
density is shown in Fig.2 where AF denotes antiferromagnetic region.
The doped holes are responsible for reducing AF correlations
which leads to an order-disorder transition.

Let us look at doped systems on the two-dimensional 
plane with respect to modulated spin structures.
Recent neutron-scattering measurements have revealed incommensurate
structures suggesting 
stripes.\cite{tra96,tra97,suz98,yama98,ara99,wak00,mat00,moo00}
The AF states with spin modulations in space have been studied 
for the one-band Hubbard model\cite{gia91,poi89,kat90,sch90,ici99} and 
t-J model\cite{whi98,whi98b,hel99} where various stripe structures are
proposed.  Our purpose is to examine the possible stripe structures and their
parameter dependence based on
the realistic three-band Hubbard model.
We can introduce a stripe in the uniform spin density state so that
doped holes occupy new levels close to the original Fermi energy keeping 
the energy loss of AF background to a minimum.

\begin{figure} 
\centerline{\psfig{figure=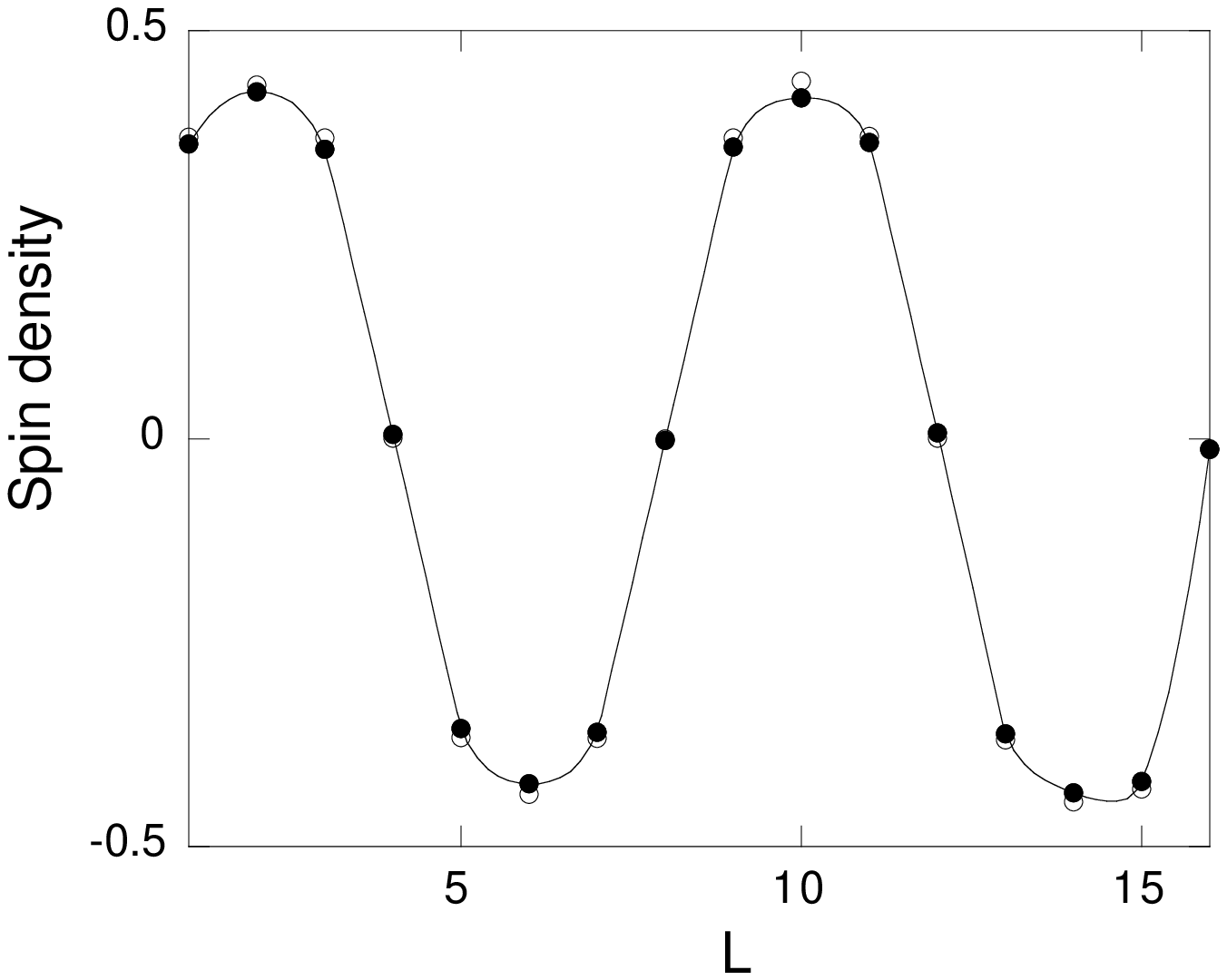,width=8cm}}
\centerline{\psfig{figure=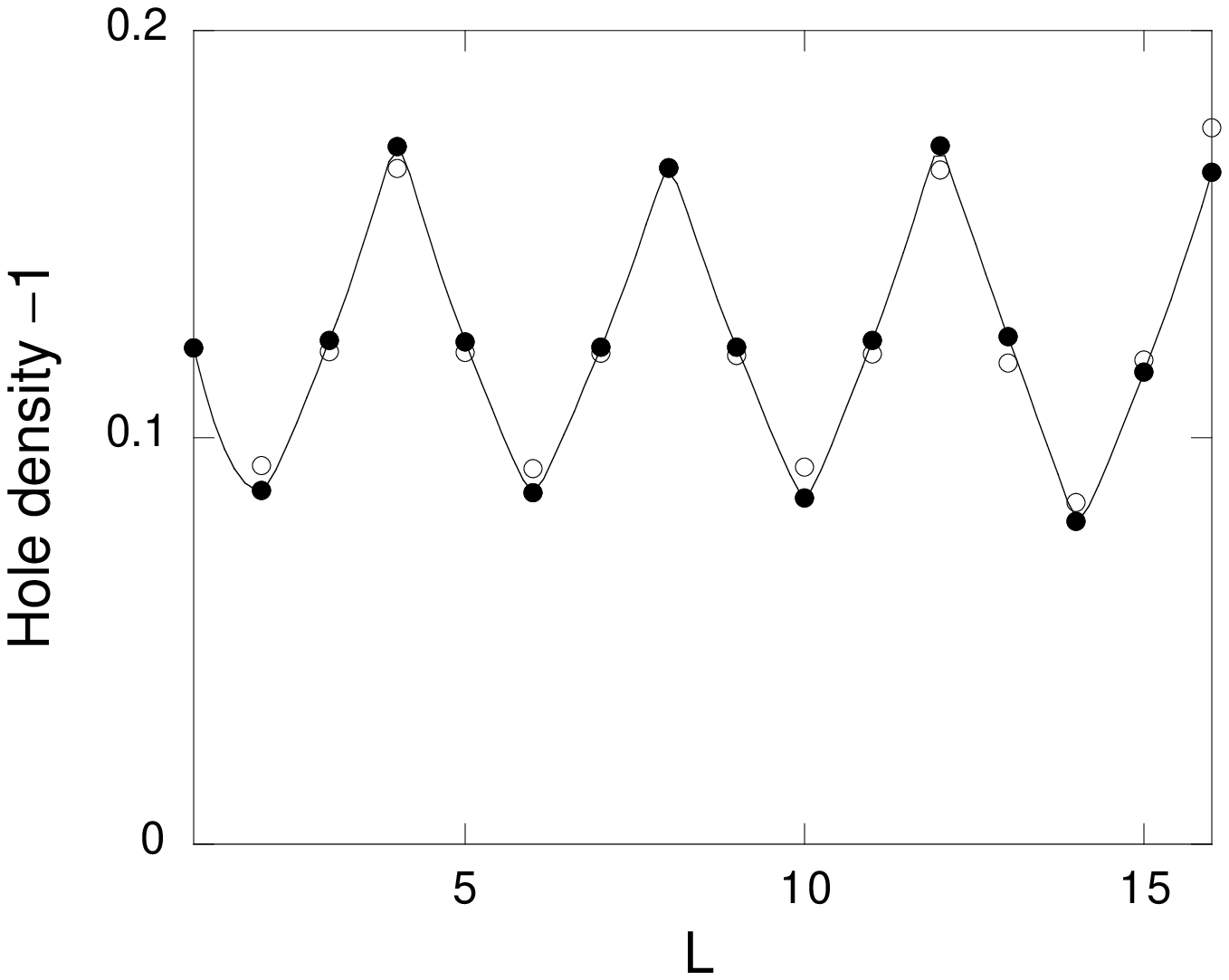,width=8cm}}
\caption{
Spin density $(-1)^{\ell-1}S_z(\ell)$ (a) and hole density (b) functions
at $\delta=1/8$ where $S_z(\ell)=n_{d\ell\uparrow}-n_{d\ell\downarrow}$.
Solid symbols are for the $16\times 16$ square lattice and open symbols are
for the $16\times 4$ rectangular lattice.
The boundary conditions are antiperiodic in $x$-direction and
periodic in $y$-direction, respectively.
}
\label{fig4}
\end{figure}

\begin{figure} 
\centerline{\psfig{figure=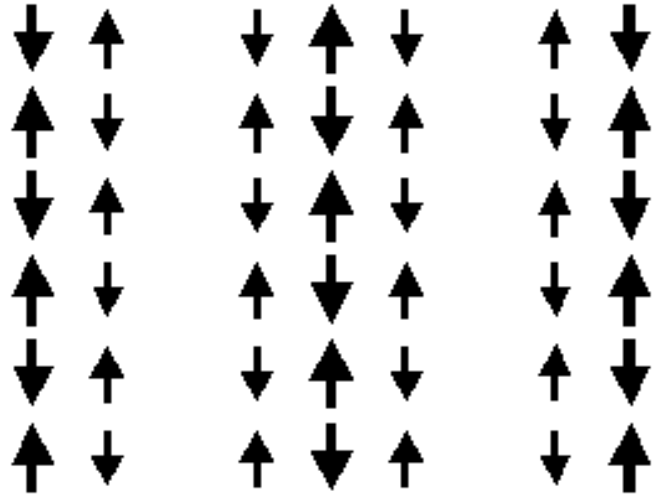,width=\columnwidth}}
\caption{
Spin structure in the incommensurate stripe state at $\delta=1/8$.
The boundary conditions are the same as in Fig.4.
}
\label{fig5}
\end{figure} 

\begin{figure}
\centerline{\psfig{figure=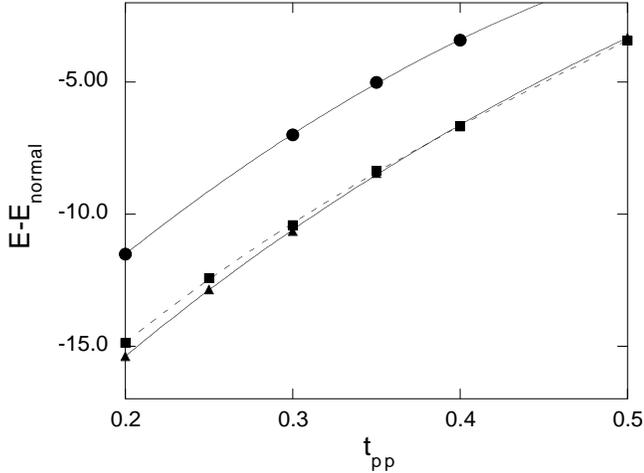,width=\columnwidth}}
\caption{
Energy per site in reference to the normal state as a function of $t_{pp}$ for 
a $16\times 16$ square lattice at $\delta=1/16$.
Circles, triangles and squares denote the energy for 4-lattice stripes,
8-lattice stripes, and commensurate SDW, respectively.
For solid symbols the boundary conditions are antiperiodic in $x$-direction and
periodic in $y$-direction, and for open triangles they are periodic in 
$x$-direction and antiperiodic in $y$-direction, respectively.
Monte Carlo statistical errors are smaller than the size of symbols.
}
\label{fig6}
\end{figure}

\begin{figure}
\centerline{\psfig{figure=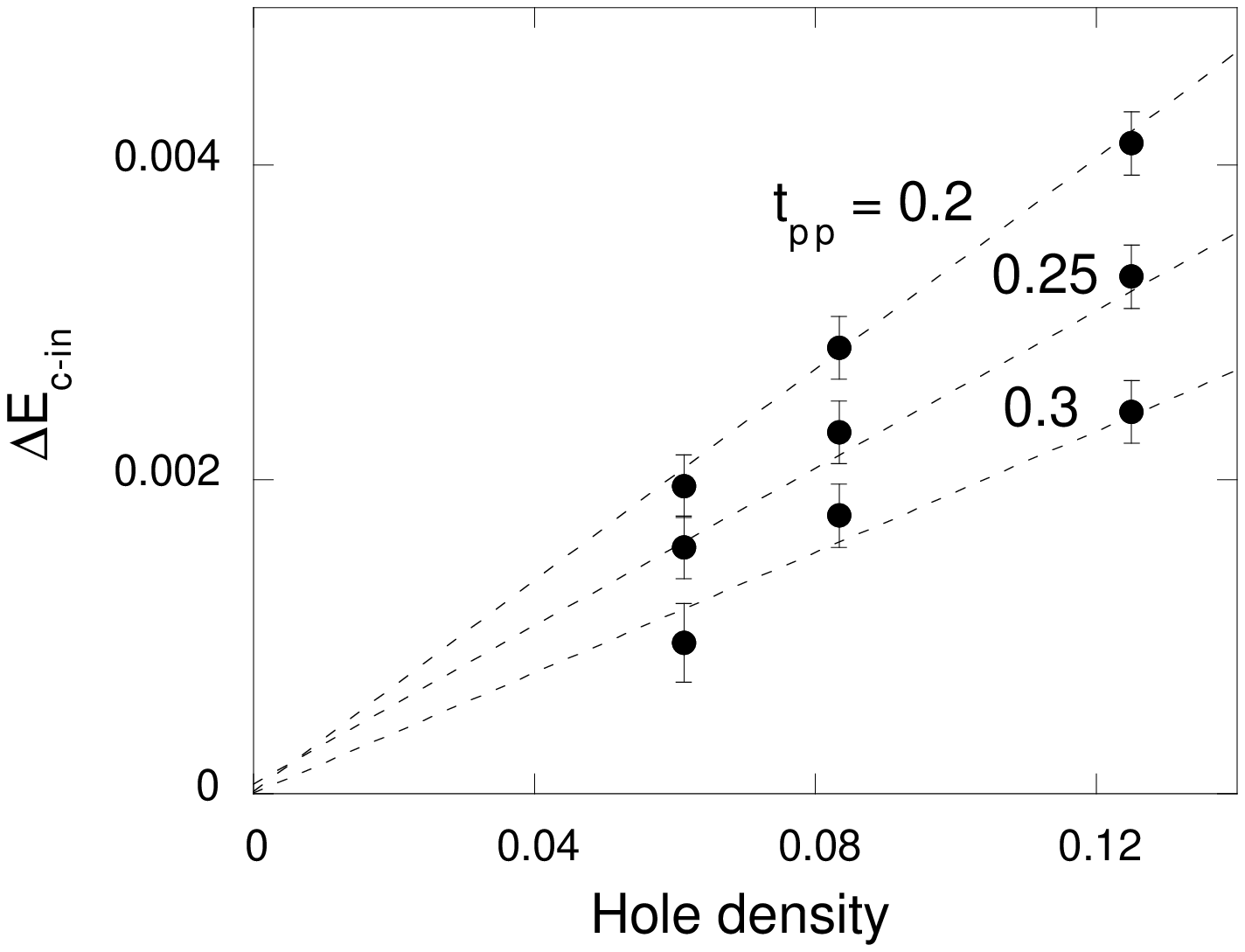,width=\columnwidth}}
\caption{
Energy difference between the commensurate and incommensurate states
at $\delta=1/16$ ($16\times 16$ lattice), $\delta=1/12$ ($24\times 12$ lattice)
and $\delta=1/8$ ($16\times 16$ lattice).
From the top
$t_{pp}=0.2$, $t_{pp}=0.25$ and $t_{pp}=0.3$.
The boundary conditions are periodic in $x$-direction and
antiperiodic in $y$-direction, respectively.
}
\label{fig7}
\end{figure}

In the actual calculations we set $\xi_c=1$ and $\xi_s=1$ in eqs.(\ref{ndi})
and (\ref{mi}) since the expected values are mostly independent of
$\xi_c$ and $\xi_s$.  We optimize $\alpha$ in eq.(\ref{ndi}) instead of
fixing it in order to lower the expected energy value further because
any eigenfunction of $H_{trial}$ can be a variational wave function.
It is also possible to assume that $\delta n_{di}$ and $m_i$ oscillate
according to the cosine curves cos($4\pi\delta x_i$) and
cos($2\pi\delta x_i$), respectively, where $\delta$ is the doping ratio.
Both methods give almost the same results within Monte Carlo statistical
errors.
Let us define $n$-lattice stripe as an incommensurate state with one
stripe per $n$ ladders for which the incommensurate wave vector is given by
$Q_s=(\pi\pm\pi/n,\pi)$ and $Q_c=(\pm 2\pi/n,0)$ for the spin and charge parts,
respectively.  The incommensurate state predicted by neutron
experiments at $\delta=1/8$ is four-lattice stripe for which
$Q_s=(\pi\pm \pi/4,\pi)$ and $Q_c=(\pm\pi/2,0)$.
In Fig.3 we show the energy for commensurate and incommensurate SDW
states on the $16\times 16$ lattice at the doping ratio $\delta$=1/8, 
where the incommensurability
is given by $\pi/4 (=2\pi\delta)$ for four-lattice stripes and $\pi/8$ for
eight-lattice stripes, respectively.
The four-lattice stripe is stable in the range of
$0.2\le t_{pp}\le 0.4$.
In Fig.3 we have shown the energy for two types of boundary conditions, which
indicates that the effect of boundary conditions is not crucial for the
$16\times 16$ system, whilst the boundary conditions change the ground
state completely for small systems such as a $16\times 4$ lattice.
The spin-correlation function exhibits an incommensurate structure as shown
in Fig.4 and the hole-density function oscillates 
corresponding to a formation of stripes.
The spin structures are illustrated in Fig.5.
The energy at $\delta=1/16$ is shown in Fig.6 where the four-lattice stripe
state has a higher energy level than for eight-lattice stripe for all values
of $t_{pp}$.
The energy gain of the incommensurate state per site in reference to the
uniform AF state denoted as $\Delta E_{c-in}$ is shown 
in Fig.7 for
$t_{pp}=0.2$, 0.25 and 0.3.

The incommensurability $\Delta q/(2\pi)$ for $t_{pp}=0.3$ is also shown in 
Fig.8 by solid circles, which is proportional to the doping ratio and is
consistent with the neutron-scattering experiments for
incommensurability.\cite{yama98} 
This should be compared with the variational Monte Carlo evaluations for 
the one-band Hubbard
model\cite{gia91} where the stripe states with large intervals are shown
to be stable.  Inorder to explain the linear dependence of $\Delta q/(2\pi)$
on the hole density, the effect of $t_{pp}$ should be taken into account.
The energy gain due to a formation of stripes is approximately proportional
to the number of stripes.
The size dependence of $\Delta E_{c-in}$ is presented in Fig.9; we
observe a tendency that $\Delta E_{c-in}$ increases as the system
size $N$ increases.  The energy gain in the bulk limit is given by
0.002$t_{dp}\approx$ 3meV for $t_{pp}=0.3$ where
$t_{dp}=1.5$eV.\cite{esk89,hyb90,mcm90}

We present typical energy scales obtained from variational Monte Carlo
calculations in terms of $t_{dp}$ in Table I.  The energy scales for 
superconductivity are consistent with 
experimental suggestions and energy difference $\Delta E_{c-in}$ between  
commensurate and incommensurate states are greater than the SC condensation
energy by one order of magnitude.  The commensurate AF energy
gain in reference to the normal state (denoted as $\Delta E_{AF}$) is larger
than $\Delta E_{AF}$ by one order of magnitude in the low-doping region.

\begin{figure}
\centerline{\psfig{figure=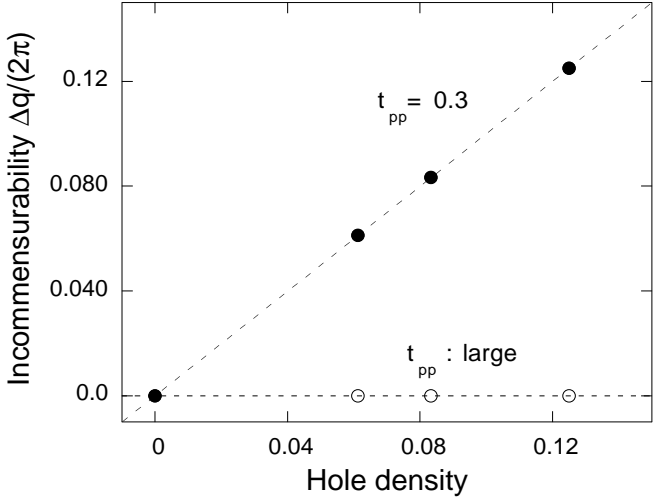,width=\columnwidth}}
\caption{
Solid circles denote incommensurability $\Delta q/(2\pi)$ for $t_{pp}=0.3$ where
the incommensurability is proportional to the hole density.
For large $t_{pp}$ values the incommensurability equals zero as shown by the open 
circles.
The boundary conditions are the same as in Fig.7.
}
\label{fig8}
\end{figure}

\begin{figure}
\centerline{\psfig{figure=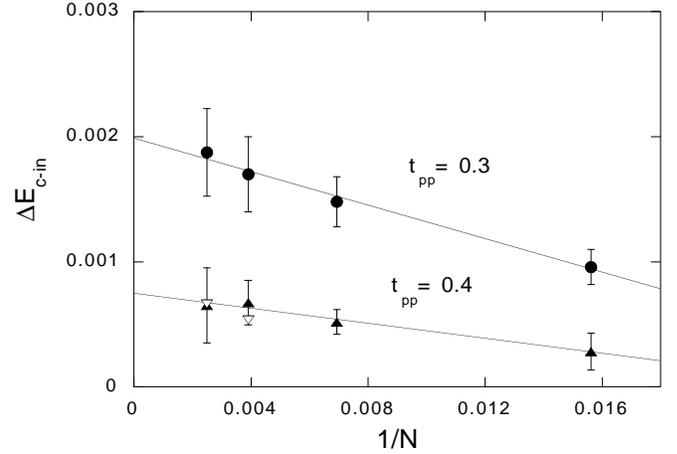,width=\columnwidth}}
\caption{
Energy difference between the commensurate and incommensurate states as
a function of $1/N$ for $t_{pp}=0.3$ (circles) and $t_{pp}=0.4$ (triangles).
Solid symbols are for rectangular lattices ($16\times 4$, $24\times 6$, 
$\cdots$), and open symbols are for square lattices ($16\times 16$, $\cdots$)
The incommensurate state is assumed to be the 4-lattice stripe state.
The boundary conditions are periodic in $x$-direction and
antiperiodic in $y$-direction, respectively.
}
\label{fig9}
\end{figure}

\onecolumn
\begin{table}
\caption{Typical energy scales obtained from variational Monte Carlo
calculations for $U=8$, $t_{pp}\approx0.3$ and $\epsilon_p-\epsilon_d=2$.
$\Delta_{AF}$ and $\Delta E_{AF}$ denote the magnitude of AF order parameter
and the AF energy gain compared to the normal state
at half-filling $\delta=0$, respectively.
$\Delta_{SC}$ and $\Delta E_{SC}$ represent the optimized SC order parameter
and SC energy gain at $\delta\sim 0.2$, respectively.
The last column indicates experimental suggestions.
}
\begin{tabular}{llll}
 \null & doping ratio &Energy($t_{dp}$)&Exp.\\
\hline
$\Delta_{SC}$ & &0.01$\sim$0.015(=15$\sim$ 20meV)& 10$\sim$ 20meV\cite{kir87,kas98}\\
$\Delta E_{SC}$ & & $\sim0.0005$(=0.75meV)\cite{yan00,yan01b}& 0.17$\sim$0.26meV\cite{lor93,and98}\\
$\Delta_{AF}$ &  $\delta=0$ & $\sim 0.6$(=900meV)& \\
$\Delta E_{AF}$ & $\delta=0$ & $\sim 0.06$(=90meV)& \\
$\Delta_{AF}$ & $\delta=1/8$ & $\sim 0.4$ & \\
$\Delta_{incom}$ & $\delta=1/8$ & $\sim 0.6$ & \\
$\Delta E_{c-in}$ & $\delta=1/8$ & $\sim 0.002$(=3meV) & 
\end{tabular}
\end{table} 
\twocolumn

\section{Summary}
We have presented our evaluations for the two-dimensional three-band Hubbard 
model using 
the variational Monte Carlo method.  We have examined an effect produced by 
holes doped into the AF state in the low-doping region.
The boundary of AF phase is dependent on $U$ as shown in the
phase diagram in Fig.2.
The inhomogeneous states with stripes are stabilized due to hole doping so 
that the energy loss of the AF background is kept to a minimum with the 
kinetic-energy gain of holes compared to uniform (commensurate) 
AF state.
In large systems the effect of boundary conditions is reduced in our
evaluations.
The distance between stripes is dependent upon the transfer integral
$t_{pp}$ between oxygen orbitals in the three-band model.
There is a region where incommensurability is proportional to
the doping ratio $\delta$ when $\delta$ is small and the energy gain due to 
a stripe formation is approximately proportional to the number of stripes.
A linearity of the incommensurability is consistent with the neutron-scattering
measurements\cite{mat00}.
It is expected that the inhomogeneity
plays an important role in the underdoped region with respect to
anomalous metallic properties in high-$T_c$ superconductors.
We have also shown the typical energy scales obtained from variational
Monte Carlo calculations.  It has been already established that the condensation 
energy
$\Delta E_{SC}$
and the magnitude of order parameter for superconductivity are in reasonable
agreement with the experimental results\cite{yan00}.  The energy gain due to 
AF ordering is larger than $\Delta E_{SC}$ by about
two order of magnitude and the energy difference between the commensurate
and incommensurate states is larger than $\Delta E_{SC}$ by one order.
The order of AF energy gain in reference to the normal state approximately
agrees with that for the t-J model\cite{yok96}.
Our evaluations seem to overestimate the antiferromagnetic energy because of
the simplicity of the Gutzwiller wave functions, which may give a
starting point for more sophisticated evaluations such as Green function
Monte Carlo approaches.

\end{document}